\documentclass[aps,pra,twocolumn,showpacs,superscriptaddress]{revtex4}

\usepackage{graphicx}
\usepackage{amsmath}
\usepackage{amssymb}
\input{epsf}
\begin{document}

\title{
Energetics 
and structural properties of three-dimensional bosonic clusters
near threshold}

 \author{G. J. Hanna and D. Blume}
\affiliation{Department of Physics and Astronomy, Washington State University,
  Pullman, Washington 99164-2814}

\date{\today}

\begin{abstract}
We treat three-dimensional bosonic clusters  with
up to $N=40$ atoms, interacting
additively through two-body Van der Waals potentials, 
in the near-threshold regime.
Our study includes super-borromean systems with $N$ atoms for which all
subsystems are unbound. We determine the energetics and structural
properties such as the expectation value of the interparticle distance
as a function of the coupling strength. 
It has been shown that the coupling strength $g_*^{(N)}$, 
for which the $N$-body
system becomes unbound, is bounded by the coupling constant $g^{(N-1)}_*$, for
which the next smaller system with $N-1$ atoms becomes unbound,
i.e., $g^{(N)}_* \ge (N-1)g^{(N-1)}_*/N$.
By fitting our numerically determined
ground state energies to a simple functional form
with three fitting parameters, 
we determine the relationship
between $g^{(N)}_*$ and $g^{(N-1)}_*$.
Our trimer and tetramer energies
fall on the so-called Tjon line, which has been
studied in nuclear physics.
We confirm the existence of
generalized Tjon lines for larger clusters.
Signatures of the universal behavior of weakly-bound three-dimensional
clusters can possibly be observed in 
ultracold Bose gases.

\end{abstract}


\maketitle

\section{Introduction}
\label{sec_introduction}
Weakly-bound few-body systems 
have been studied extensively  by the atomic, nuclear and 
condensed matter physics community
since the early days of quantum mechanics.
Within the framework of non-relativistic quantum mechanics,
the properties of a many-body
system 
are determined by 
the mass and statistics of the constituents and by 
the potential energy surface.
In many cases,
the many-body potential energy surface can be approximated
quite accurately
by a sum of two-body potentials.
The interaction strength $g$ of the $N$-body system 
is then
determined by
the underlying two-body potential and, assuming $N$ identical
mass $m$ particles,   the 
mass $m$.
The critical coupling
constant $g^{(N)}_*$, for which the $N$-body cluster becomes unbound, 
defines the threshold.
This paper investigates the near-threshold regime of bosonic 
three-dimensional $N$-body clusters,
i.e., the regime where $g \gtrsim g^{(N)}_*$.
This near-threshold regime is particularly
interesting since some properties of the bosonic many-body system
become independent of the details of the underlying potential
energy surface, i.e., some properties of weakly-bound clusters 
consisting of $N$ bosons
become universal as 
$g \rightarrow g^{(N)}_*$~\cite{bruc76,cram77,lim77,lim80,naka83,adhi88,mosz00,plat04,braa06}.

Our study includes the characterization of ``super-borromean'' $N$-body
clusters~\cite{blum02}.
Borromean trimers, which consist of three bosons 
for which each dimeric subsystem
is unbound, have been studied in detail in the 
literature~\cite{fedo94,fedo94a,jens04}.
Super-borromean clusters, 
which consist of $N$ bosons for which all
subsystems with $N-n$, where $n=1,\cdots, N-2$, are unbound,
in contrast, 
have not been studied in much detail. 
To characterize these delicate systems, we perform precise diffusion Monte
Carlo (DMC) calculations for 
clusters interacting additively through realistic shape-dependent
two-body Van der Waals potentials.
We determine the critical coupling strengths
$g_*^{(N)}$ for atomic clusters with up to $N=40$ bosons and
compare with variational bounds.
The near-threshold behavior of weakly-bound 
three-dimensional bosonic clusters
has been investigated in a series of papers 
before~\cite{cram77,lim80,mosz00,plat04}. 
We believe, however, that advances in the theoretical understanding, 
including predictions derived using effective theories and
zero-range models~\cite{plat04,yama06}, and in the numerical treatment
make it worthwhile to revisit the characterization
of weakly-bound three-dimensional clusters.
In particular, we present more accurate energies for a larger
range of coupling strengths and for a larger range of cluster sizes
than
previous studies.

The present study is additionally motivated
by recent experiments on
extremely weakly-bound
molecules created from
ultracold
Bose and Fermi gases.
Utilizing Feshbach resonances the effective
interaction strength between two atoms at ultracold
temperatures can be changed essentially at will through application of an 
external magnetic field~\cite{stwa76a,ties93}.
The existence of this external knob has led to the 
observation of extremely weakly-bound
diatomic molecules in highly-excited vibrational 
states~\cite{donl02,rega04,zwie04}
and provided evidence for the formation of Efimov trimer 
states~\cite{efim70,krae06}
in an ultracold environment.
Furthermore, recent experiments on cold Cs atoms
evidence the creation of larger weakly-bound
clusters~\cite{chin05}; 
these experiments point towards Feshbach engineering of
weakly-bound clusters.
Feshbach resonances arise from the coupling of two Born-Oppenheimer 
potential curves through 
a hyperfine Hamiltonian and require, in general, a multi-channel description. 
In the case of a broad resonance, however, the change of the effective
scattering length can be described within a single channel 
model~\cite{kohl06}.
Using a single channel approximation,
this paper describes weakly-bound three-dimensional bosonic
clusters with varying atom-atom scattering lengths with up to $N=40$ atoms.

Section~\ref{sec_system} describes the many-body
Hamiltonian and the characteristics of the underlying
two-body potential. 
Section~\ref{sec_numerical} is devoted to a discussion
of our numerical approach to solving the many-body Schr\"odinger equation.
Our results for the energetics and structural properties
are presented in Secs.~\ref{sec_results_energetics} 
and \ref{sec_results_structure},
respectively, and our conclusions in Sec.~\ref{sec_conclusion}.

\section{System and numerical approach}
\label{sec_theory}
\subsection{Many-body Hamiltonian}
\label{sec_system}
\begin{figure}
      \centering
      \includegraphics[angle=270,width=8cm]{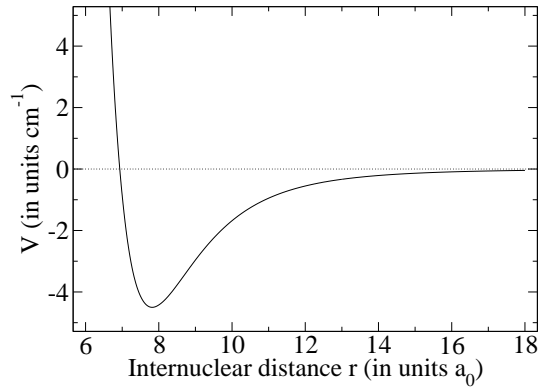}
\caption{Triplet tritium-tritium interaction
potential as a function of the internuclear distance $r$.}
\label{fig1}
\end{figure}

Consider the Hamiltonian $H$ for 
$N$ bosons with mass $m$,
\begin{eqnarray}
\label{eq_ham}
	H = -\frac{\hbar^2}{2m} \sum_{j=1}^N\nabla_j^2 + 
	\sum_{j<k}^N V(r_{jk}),
\end{eqnarray}
where $\nabla_j^2$ and $r_{jk}$ denote respectively
the 3D Laplace operator of the $j$th boson and 
the internuclear distance between particles $j$ and $k$.
This Hamiltonian assumes a many-body potential energy surface
written as a sum of atom-atom potentials $V(r)$.
Our calculations are performed for 
a realistic Van der Waals
triplet tritium-tritium 
potential~\cite{kolo74,kolo90,jami00,bukt74,herr64,yan96}, 
which is repulsive at
short interparticle distances $r$ and falls
off at large $r$ as $\sum_{n=6,8,\cdots}-C_n r^{-n}$.
Figure~\ref{fig1} shows the tritium-tritium potential
as a function of the interparticle distance $r$.
The potential has a minimum of depth $-4.6$cm$^{-1}$
at $r \approx 7.8a_0$, where $a_0$ denotes the Bohr radius. 
Solving the one-dimensional scaled radial
Schr\"odinger equation shows that the tritium dimer
has no bound state~\cite{blum02}
(see also Sec.~\ref{sec_dimer}).
The scattering length $a$,
\begin{eqnarray}
\label{eq_scatt}
a = \lim_{k \rightarrow 0} -\frac{\tan(\delta(k))}{k},
\end{eqnarray}
of the tritium dimer is negative, i.e., $a=-82.1a_0$~\cite{blum02}, which
indicates that the
dimer is only slightly short of binding. 
In Eq.~(\ref{eq_scatt}), $\delta(k)$ denotes the energy-dependent
$s$-wave phase shift and $k$ the relative 
wave vector of the equivalent one-body
problem with reduced mass $m/2$.

Our interest in this paper is in a detailed description of
weakly-bound clusters with varying coupling constant near threshold.
To change the coupling strength $g$ of
the cluster, we vary the atom mass $m$, i.e.,
we consider ``artificial'' clusters with 
atom masses that are heavier and lighter than the tritium mass.
By rewriting the many-body Schr\"odinger equation in scaled units,
it can be readily seen that
changing the atom mass changes the coupling strength.
For example, for systems interacting additively through Lenard-Jones potentials
with well depth $\epsilon$ and length scale
$\sigma$
the coupling constant $g$ is directly
proportional to the atom mass, $g=4m \epsilon \sigma^2 / \hbar^2$.

As alluded to in the introduction, the coupling strength can 
be varied experimentally via a Feshbach resonance~\cite{cour98,inou98}.
Although a full description of a
Feshbach resonance requires the coupling between at least two 
channels--- in tritium, e.g., of the singlet and 
triplet potential curves
(coupled through a long-range hyperfine 
Hamiltonian)~\cite{burk98,blum02}---, some
properties can be described within a single channel model. We thus 
envision that changing the atom mass in our single channel treatment can
be mapped to changing the strength of an external magnetic field,
and hence of the
atom-atom scattering length, 
in the vicinity of a two-body Feshbach resonance.
We expect that our calculations uncover the ``generic''
behaviors of three-dimensional bosonic Van der Waals clusters, 
which are interacting additively through two-body potentials with repulsive
short-range core
and long-range tail with leading $-C_6/r^6$ term.
In particular,
we believe that usage of a different two-body potential 
in Eq.~(\ref{eq_ham}) will result in the same
qualitative but possibly different quantitative behaviors
of weakly-bound bosonic clusters.

\begin{figure}
      \centering
      \includegraphics[angle=270,width=8cm]{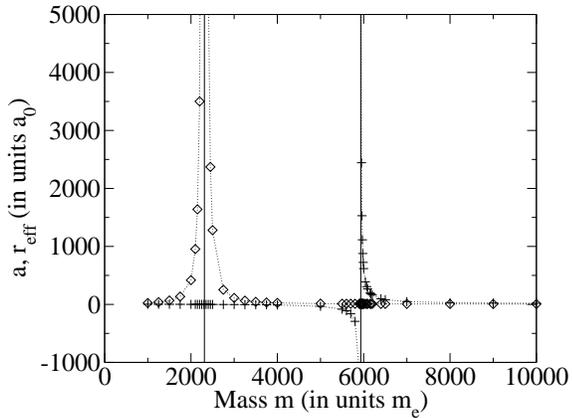}
\caption{Pluses and diamonds show the atom-atom 
scattering length $a$ and the effective range $r_{{eff}}$, respectively,
as a function of the atom mass $m$
for the triplet tritium-tritium
potential (see text). To guide the eye dotted lines connect the symbols.
Vertical solid
lines indicate the mass $m \approx 2311.0 m_e$, 
at which $a$ 
goes through zero, and the critical mass $m^{(2)}_* \approx 5933.4 m_e$, 
at which
the scattering length $a$ diverges.
}
\label{fig2}
\end{figure}

Pluses in Fig.~\ref{fig2} show the atom-atom scattering length $a$
for the tritium-tritium potential
as a function of the atom mass $m$.
The scattering length $a$ diverges at $m \approx 5933.4(2) m_e$,
where $m_e$ denotes the electron mass
and the value in round brackets the uncertainty of $m$
arising from the numerical determination of the scattering length $a$.
Since this is the mass
at which the dimer becomes unbound, we refer to this mass as the critical 
mass $m^{(2)}_*$ of the dimer. A vertical solid line
in Fig.~\ref{fig2} marks the value of $m_*^{(2)}$. We find that
the scattering length $a$ vanishes for $m \approx 2311.0(2) m_e$. 
This mass value is indicated by a vertical solid line in Fig.~\ref{fig2}
and puts an upper bound on the critical mass for the bulk system 
($N \rightarrow \infty$), i.e.,
$m^{(\infty)}_* \le  2311.0 m_e$~\cite{bruc76}. This bound is
obtained variationally by expanding the energy in terms of the density. 
For negative $a$, the leading order in the expansion becomes negative
and the bulk system is necessarily bound~\cite{bruc76}.
Section~\ref{sec_cluster} 
compares our critical masses $m^{(N)}_*$ 
calculated for up to $N=40$ atoms 
with the variational upper bound for
$m^{(\infty)}_*$.

Diamonds in
Fig.~\ref{fig2} show the effective range $r_{{eff}}$,
which we calculate through the relationship
\begin{eqnarray}
\label{eq_reff}
-\frac{1}{a(k)} \approx 
-\frac{1}{a}+ \frac{1}{2} r_{{eff}} k^2,
\end{eqnarray}
as a function of the atom mass $m$. 
In Eq.~(\ref{eq_reff}), 
$a(k)$ denotes the energy-dependent scattering length,
$a(k)=-\tan(\delta(k))/k$.
The effective range is
largest in the region
where the two-body scattering length vanishes. In the region where 
$a$ diverges, $r_{{eff}}$ takes on values of the order of 
$10a_0$. 
The scattering length $a$, the
effective range $r_{{eff}}$ and
the Van der Waals length $r_{{VdW}}$,
where $r_{{VdW}}=(m C_6/\hbar^2)^{1/4}$, are
the relevant length scales of the two-body problem near threshold.
For a zero-range model with a single parameter,
namely the scattering length $a$,
to be applicable for $N=2$, 
$a$ needs to be the largest length scale in the problem.
This condition can be expressed as 
\begin{eqnarray}
\label{eq_cond}
|E_2| \ll \min\left(\frac{\hbar^2}{m r_{{eff}}^2},
\frac{\hbar^2}{m r_{{VdW}}^2} \right),
\end{eqnarray}
where $E_2$ denotes the ground state energy of the dimer.
For zero-range models to be applicable to clusters with $N>2$,
a condition similar to Eq.~(\ref{eq_cond}), 
possibly with an additional scaling factor $N$ or
$N(N-1)/2$, needs to be fulfilled.

\subsection{Numerical treatment of $N$-body clusters}
\label{sec_numerical}
To determine the ground state energy and wave function
of the two-body system
[Eq.~(\ref{eq_ham}) with $N=2$],
we separate off the center of mass motion and scale the
wave function for the interparticle distance
to remove first derivative terms in the kinetic energy operator.
The scaled one-dimensional radial Schr\"odinger equation 
can then be solved by diagonalizing the Hamiltonian using B-splines.
To treat very weakly-bound dimers with varying mass $m$,
we optimize the adaptive grid (i.e., the
grid spacings, the number of grid points and the integration interval)
for each mass. The upper integration limit is determined by the
size of the bound state; integrating the Schr\"odinger equation out
to roughly $100 a$ leads to converged results for all
two-body systems considered in Sec.~\ref{sec_dimer}.

The calculations of the trimer energies are, due to the larger number
of degrees of freedom, more involved
than those of the dimer energies. Separating off the center of mass 
motion reduces the nine-dimensional problem to a six-dimensional
problem. Since we are in this paper
primarily interested in ground state properties, we restrict
ourselves to states with vanishing total angular momentum, i.e., $J=0$.
The resulting three-dimensional
Hamiltonian can be written in terms of Whitten and Smith' 
hyperspherical 
coordinates~\cite{whit68}, which allow 
the Bose symmetry to be accounted for readily. 
To solve the
corresponding scaled Schr\"odinger equation, we 
expand the wave function in angle-dependent channel functions
$\Phi$, which depend  parametrically on the hyperradius $R$,
and a set of weight functions $F(R)$. 
Our numerical implemention is described
in
Ref.~\cite{blum00a}.
We check the convergence by changing the hyperangular 
grid, the hyperradial grid, the number of channel functions
included in the expansion, and the
step size used in the numerical determination of the derivatives of the
channel functions. 
The trimer ground state energies presented in Sec.~\ref{sec_trimer} 
have an accuracy of a few percent.
For selected trimers, we also report the first
excited state energy with $J=0$.

For cluster systems with more than a few atoms, 
basis set expansion-type techniques
become computationally
unfeasible. Consequently, we
solve the many-body Schr\"odinger equation for $N \ge 4$ using
alternative techniques, i.e., the variational quantum Monte Carlo (VMC)
method and the DMC method with importance sampling~\cite{hamm94}.
Our numerical implementations follow Ref.~\cite{blum96}.
The VMC method minimizes the energy of the cluster
system by optimizing the many-body wave function, which is written
in terms of a set of parameters $\vec{p}$. 
The optimized variational wave function $\psi_T$
then enters
our DMC calculations, which result in
essentially exact ground state energies, as a guiding function. 
Due to the stochastic nature of the MC algorithms,
the DMC energies reported in Sec.~\ref{sec_cluster}
have statistical uncertainties. 

We use two different functional forms for the 
variational wave function. For small clusters with about
up to $N=10$ atoms, each atom
has roughly the same
average distance to all other atoms in the cluster.
In this case, our variational wave function $\psi_T$ 
is written as a product
of pair wave functions $\phi$~\cite{lewe97},
\begin{eqnarray}
\label{eq_trial0}
\psi_T(\vec{r}_1, \cdots, \vec{r}_N) = \prod_{j<k}^N \phi(r_{jk}),
\end{eqnarray}
where
\begin{eqnarray}
\label{eq_trial1}
\phi(r) = \exp \left[ -\frac{p_{-5}}{r^{5}} - \frac{p_{2}}{r^{-2}} -
p_0 \log(r) - p_1 r 
\right].
\end{eqnarray}
For larger clusters, the variational wave function given in 
Eqs.~(\ref{eq_trial0}) and (\ref{eq_trial1})
does not give a good variational energy
and we additionally
include a variational Fermi function which depends on the
distance $R_{j}$ of the $j$th atom to the center of mass of the 
cluster~\cite{rama90},
\begin{eqnarray}
\label{eq_trial0a}
\psi_T(\vec{r}_1, \cdots, \vec{r}_N) = \left[ \prod_{j<k}^N \phi(r_{jk})
\right]
\left[
\prod_{l=1}^N \bar{\phi}(R_{l}) \right].
\end{eqnarray}
Here,
$\phi$ is given by Eq.~(\ref{eq_trial1}) with $p_0=p_1=0$ and
\begin{eqnarray}
\label{eq_trial2}
\bar{\phi}(R) = \left[ 1 + \exp \left( \frac{R - p_e}{p_{\sigma}}
\right) \right] ^{-1}.
\end{eqnarray}
The variational
parameters $p_e$ and $p_{\sigma}$
determine the size of the cluster
and  the sharpness of the cluster's surface region, respectively.
For each cluster system considered, we optimize 
the variational parameters
by minimizing the
energy expectation value
$\langle \psi_T | H | \psi_T \rangle / \langle \psi_T | \psi_T \rangle$.
The $3N$-dimensional integrals are evaluated 
using the Metropolis algorithm.
Our VMC energies, except for those for the systems closest to threshold
(see Sec.~\ref{sec_cluster}),
recover more than about 75-80\% 
of the essentially exact DMC ground state energies. 

The DMC calculations become computationally more demanding 
as we approach the threshold,
since the kinetic and potential energy nearly cancel. In the
near-threshold regime great
care has to be taken to avoid any guiding function bias and to
ensure convergence of the DMC calculations.
To check that our DMC code describes extremely weakly-bound clusters
accurately, 
we compare the DMC energies for the trimer 
with those calculated by the hyperspherical B-spline treatment.
We find agreement to within the statistical uncertainty
for $m \ge 6000 m_e$ but do not obtain reliable DMC energies
for significantly smaller masses.

Unlike the DMC energy expectation value, which is 
essentially exact (except for statistical uncertainties and possible
time step errors), the expectation value
of any structural quantity $B$ 
is in the ``standard'' DMC algorithm calculated with
respect to the mixed density,
\begin{eqnarray}
\langle B \rangle_{DMC} = 
\langle \Psi | B | \psi_T \rangle /
\langle \Psi | \psi_T \rangle.
\end{eqnarray}
Here,
$\Psi$ denotes the exact stationary ground state wave function~\cite{hamm94}.
To improve upon this mixed estimator, we 
calculate the so-called extrapolated expectation value 
$\langle B \rangle_{ex}$~\cite{whit79},
\begin{eqnarray}
\label{eq_extra}
\langle B \rangle_{ex} = 
2 \langle B \rangle_{DMC} - \langle B \rangle_{VMC},
\end{eqnarray}
where $ \langle B \rangle_{VMC} $
denotes 
the VMC expectation value, 
\begin{eqnarray}
\langle B \rangle_{VMC} = 
\langle \psi_T | B | \psi_T \rangle /
\langle \psi_T | \psi_T \rangle.
\end{eqnarray}
For the systems studied in this paper,
the extrapolated expectation values
$\langle B \rangle_{ex}$ are expected to be
fairly close to the exact expectation values.
Section~\ref{sec_results_structure} reports 
expectation values for
the pair
distribution function $P(r)$ and 
the interparticle distance $r$.

\section{Energetics}
\label{sec_results_energetics}
This section presents our numerically determined energies 
for clusters with up to 40 atoms and their
interpretation.

\subsection{$N=2$}
\label{sec_dimer}
\begin{figure}
      \centering
      \includegraphics[angle=0,width=8cm]{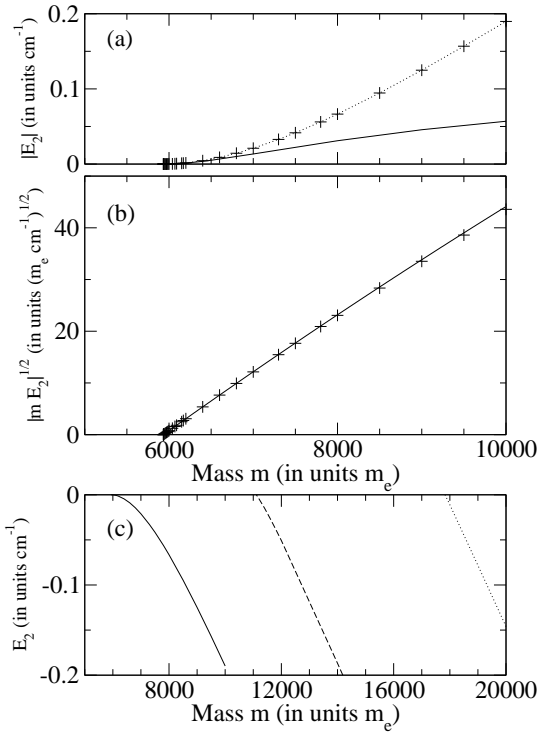}
\caption{
(a) Pluses show the absolute value of the numerically determined
two-body ground state 
energies $E_2$ as a function of the atom mass $m$.
To guide the eye, a dotted line connects the symbols.
For comparison, a solid line shows the absolute value
of the two-body energies $E_2^{\delta}$, Eq.~(\ref{eq_delta}), obtained
from the scattering length $a$.
(b) Pluses show the scaled dimer ground state energies $\sqrt{m |E_2|}$ 
as a function of the atom mass $m$.
The scaled energies $\sqrt{m |E_2|}$ vary to first order linearly with $m$.
A solid line shows our fit to the scaled energies 
$\sqrt{m |E_2|}$, 
treating $c_1^{(2)}$, $c_2^{(2)}$ and $m^{(2)}_*$ as fitting parameters
(see text).
(c) Solid, dashed and dotted lines show the lowest two-body energies
for $l=0$, $1$ and $2$, respectively, as a function of the
atom mass $m$.
Note that the mass range shown in the lowest panel differs
from the mass ranges shown in the upper two panels.
}
\label{fig3}
\end{figure}

Pluses in Fig.~\ref{fig3}(a) show the absolute value of the
$s$-wave ground state energies $E_2$ for two
particles
interacting through the triplet tritium-tritium potential 
for a number of different $m$, i.e., $m \in [5933.4m_e ,10000m_e]$.
The tritium dimer itself is, as mentioned in Sec.~\ref{sec_system}, 
unbound. 
The ground state energies shown in Fig.~\ref{fig3}
extend over nearly ten orders of magnitude;
$E_2$ for the most weakly-bound dimer
considered with $m=5933.4$ is $-6 \times 10^{-11}$cm$^{-11}$, 
and that for the most strongly-bound
dimer considered with $m=10000 m_e$ is $-0.19$cm$^{-1}$.

We can compare the numerically determined ground state energies $E_2$
with the energies $E_2^{\delta}$ predicted
from a
zero-range model, which supports a bound state for positive $a$,
\begin{eqnarray}
\label{eq_delta}
E_2^{\delta}= -\frac{\hbar^2}{m a^2} .
\end{eqnarray}
A solid line in Fig.~\ref{fig3}(a) 
shows the absolute value of the zero-range energies $E_2^{\delta}$.
In the region where the zero-range model
provides a good description of the dimer energies, the 
scattering length is the largest length scale in the problem, i.e.,
$a \gg r_{eff}$ and $a \gg r_{VdW}$
(see Fig.~\ref{fig2}). 
As $a$ becomes comparable to $r_{{eff}}$, 
the energies $E_2^{\delta}$ deviate visibly 
from the exact energies $E_2$
and Eq.~(\ref{eq_delta}) has to be modified 
to account for the dependence of the energy
on the effective range $r_{eff}$
in addition to $a$.

We now describe an analysis that allows an accurate 
determination  of 
the critical mass $m_*^{(2)}$ from the two-body ground state energies. 
In principle, this analysis is not needed 
since our scattering length calculations allow the critical
mass $m^{(2)}_*$ to be determined 
with high accuracy (see Sec.~\ref{sec_system}).
The analysis presented in the next two paragraphs for $N=2$ is 
meant as a test
of principle; an analogous analysis is in Secs.~\ref{sec_trimer} 
and \ref{sec_cluster} applied to larger clusters.
For $N=3$, accurate 
calculations of the dimer plus atom
scattering length can be performed but are not pursued here.
For $N > 3$, however, only 
approximate 
calculations for the $N-1$ plus atom scattering length
have been performed to date~\cite{naka83,blum00}.

Within effective range theory, the two-body ground state energies $E_2$ near
threshold are determined by Taylor-expanding the logarithmic derivative
of the bound state wave function about the
critical mass $m_*^{(2)}$~\cite{blatt,newton},
\begin{eqnarray}
\label{eq_expand}
\sqrt{m |E_2|}= \sum_{i=1}^{\infty} c_i^{(2)} (m-m_*^{(2)})^i,
\end{eqnarray}
where the $c_i^{(2)}$ denote expansion coefficients.
Pluses in Fig.~\ref{fig3}(b) show the scaled energies
$\sqrt{m |E_2|}$,
which
vary roughly linearly with $m$.
Close inspection, however,
reveals deviations from a linear behavior.
This indicates that the first term in the expansion 
given by Eq.~(\ref{eq_expand}) is dominant,
but that the second expansion coefficient $c_2^{(2)}$ 
contributes non-neglegibly.
To determine $m_*^{(2)}$, $c_1^{(2)}$
and $c_2^{(2)}$, we fit our scaled energies
for $m \in [5933.4m_e,10000m_e]$ to the first two terms of the
expansion given by Eq.~(\ref{eq_expand}). The resulting fit with
$m^{(2)}_*=5933.5(1)m_e$, 
$c_1^{(2)}=1.1623(2) \times 10^{-2} \sqrt{\mbox{cm}^{-1}/m_e}$ and 
$c_2^{(2)}= -2.243(5) \times 10^{-7} \sqrt{\mbox{cm}^{-1}/m_e^3}$ 
(see Table~\ref{tab1}) agrees well with the exact energies 
and is shown by a solid line in Fig.~\ref{fig3}(b).
\begin{table}
\begin{tabular}{l|l|l|l} 
$N$ & $m_*^{(N)}$ [$m_e$] & $c_1^{(N)}$ [$\sqrt{\mbox{cm}^{-1}/m_e}$] & $c_2^{(N)}$ [$\sqrt{\mbox{cm}^{-1}/ m_e^3}$] \\ \hline
2  	& $5933.5(1)$ &	$0.011623(2)$ & $-2.243(5) \times 10^{-7}$ \\
3  	& $5352(17)$  &	$0.026(2)$ & $-2.6(6) \times 10^{-6}$ \\
4  	& $4836(08)$  &	$0.033(1)$ & $-1.9(2) \times 10^{-6}$ \\
5  	& $4527(15)$  &	$0.042(2)$ & $-3.2(6) \times 10^{-6}$\\
6  	& $4311(18)$  &	$0.051(2)$ & $-4.2(9) \times 10^{-6}$ \\
7  	& $4097(06)$  &	$0.056(1)$ & $-3.7(4) \times 10^{-6}$ \\
8  	& $3992(09)$  &	$0.067(2)$ & $-7.0(9) \times 10^{-6}$ \\
9  	& $3867(16)$  &	$0.072(3)$ & $-6.6(1.7) \times 10^{-6}$ \\
10  	& $3789(14)$  &	$0.084(4)$ & $-1.0(3) \times 10^{-5}$ \\
20  	& $3142(26)$  &	$0.090(8)$ & $-1.4(6) \times 10^{-5}$ \\
40  	& $2919(13)$  &	$0.202(7)$ & $-2.9(6) \times 10^{-5}$  
\end{tabular}
\caption{Fitting parameters $m_*^{(N)}$, $c_1^{(N)}$ and $c_2^{(N)}$
for three-dimensional bosonic clusters with $N=2-10$, 20 and 40
interacting additively through
a triplet tritium-tritium potential.
The numbers in brackets indicate the uncertainties of the fit,
neglecting possible uncertainties of the energies (see text).}
\label{tab1}
\end{table}
The numbers in round brackets indicate the uncertainty of the fitting
parameters,
excluding possible numerical inaccuracies of the two-body energies.
The critical mass extracted by fitting
to the dimer bound state energies is in excellent agreement 
with the critical mass $m^{(2)}_*=5933.4(2)m_e$ 
determined from the scattering length
calculations (see Sec.~\ref{sec_system}).

The calculations for larger clusters 
necessarily cover, due to numerical difficulties, a smaller range 
of energies, i.e., we are not able to perform bound state
calculations as close to threshold as for the dimer. Furthermore,
our cluster energies for $N>3$ can only be determined within a 
statistical uncertainty, which adds an additional complication.
If we exclude the two-body energies very close to threshold from our 
fit, 
i.e., if we perform a fit to the scaled 
energies with $m \in [6800m_e,10000m_e]$ (which
is roughly
comparable to the corresponding ranges considered for $N \ge 3$, see 
Secs.~\ref{sec_trimer} and \ref{sec_cluster}), we find
a critical mass $m^{(2)}_*=5932.1(8)m_e$, where the
uncertainty in brackets reflects, as above, the uncertainty of the
fit. 
Since $c_2^{(2)}$ is negative
the critical mass predicted by this fit,
which excludes the energies closest to threshold, is expected to be smaller
than the exact threshold value.
The deviation from the 
fit that includes the whole mass range (see above) is about $1.5m_e$,
thus providing us with an estimate of the 
error made when extracting the critical mass from a set of energies,
which excludes the very near-threshold regime.

None of the dimers considered in this subsection supports 
an excited $l=0$ state, where $l$ denotes
the orbital angular momentum quantum number. Some of the dimers do, however,
support rotationally excited states with $l>0$. 
Even $l$ states are allowed for bosons and fermions with opposite spin, and
odd $l$ states for spin-aligned fermions. Figure~\ref{fig3}(c)
shows the two-body bound state energies for $l=0$ (solid line),
$l=1$ (dashed line) and $l=2$ (dotted line)
as a function of the atom mass $m$.
The near-threshold
behavior of the $l>0$ states is,
due to the presence of the angular momentum barrier of the effective 
potential,
distinctly different from that of the $l=0$ states (for which
the angular momentum barrier vanishes).
The next subsection discusses the energetics of the trimer.

\subsection{$N=3$}
\label{sec_trimer}
\begin{figure}
      \centering
      \includegraphics[angle=270,width=8cm]{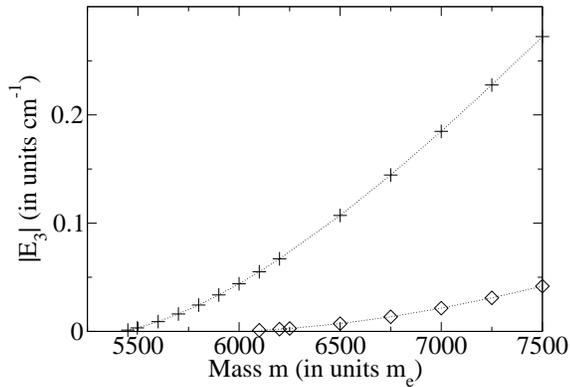}
\caption{
Pluses and diamonds show respectively 
the absolute value of the 
ground state and first excited state energies with $J=0$
of the trimer as a function of the atom mass $m$. 
To guide the eye, dotted
lines
connect the symbols.
}
\label{figtrimer}
\end{figure}

Pluses in
Fig.~\ref{figtrimer}
show the absolute value of the 
trimer ground state energies, obtained by solving
the Schr\"odinger equation
using hyperspherical coordinates (see Sec.~\ref{sec_numerical}), 
as a function of the atom mass $m$.
The ground state energies $E_3$
extend over nearly three orders of magnitude.
The most weakly-bound trimer considered 
with $m=5430 m_e$ has a ground state
energy of $-4.8 \times 10^{-4}$cm$^{-1}$, and the most strongly-bound trimer
considered with $m=7500m_e$ has a ground state energy of $-0.27$cm$^{-1}$.

As in the dimer case, we fit
the scaled energies $\sqrt{m |E_3|}$
to the expansion given in Eq.~(\ref{eq_expand})
with $c_i^{(2)}$ ($i \le 2$)
and $m_*^{(2)}$ replaced by $c_i^{(3)}$ and $m_*^{(3)}$, respectively.
The resulting fitting parameters 
$c_1^{(3)}$,
$c_2^{(3)}$ and $m_*^{(3)}$ are given in Table~\ref{tab1}.
The fit, shown by a solid line in Fig.~\ref{fig4}(b),
describes the trimer energies quite well. 
Since the trimer contains three ``dimer bonds'',
the critical mass $m_*^{(3)}$ for the trimer is significantly smaller
than that for the dimer.
The 
parameter $c_2^{(N)}$, which quantifies the 
non-linear dependence of the scaled energies on 
the mass $m$, is about an order of magnitude
more negative for the trimer than for the dimer. The negative value 
of $c_2^{(3)}$ 
suggests that the
critical mass $m_*^{(3)}$ predicted by our fit is somewhat smaller
than the true
threshold value, which could be determined
more precisely if we were able to calculate
accurate energies closer to threshold.
The smallest mass
for which we reliably determine a negative trimer energy
provides an upper bound for the critical mass $m_*^{(3)}$.
We believe that the lower bound, i.e., the critical
mass predicted by our fit, is more accurate than this
upper bound.

Diamonds in Fig.~\ref{figtrimer} show the first excited state energy
$E_3^{(1)}$ with $J=0$ as a function of $m$. 
Although the
mass at which the excited state becomes unbound is larger than that
at which the ground state becomes unbound,
the excited state energies approach the threshold in qualitatively
the same way as the ground state energies do.
In fact, 
the excited state of the
helium trimer, which is an Efimov state~\cite{lim77},
was for a long time considered to be the possibly most 
promising candidate for observing 
universal behaviors experimentally~\cite{brue05}.
Just as the near-threshold behavior of the excited dimer states
with $l>0$ 
is qualitatively different from that of the $l=0$ state
[see Fig.~\ref{fig3}(c)],
the near-threshold behavior of trimer states with $J \ne 0$ is predicted
to be qualitatively different from that of the $J=0$ states~\cite{esry01}.

\subsection{$N \le 40$}
\label{sec_cluster}
\begin{figure}
      \centering
      \includegraphics[angle=0,width=8cm]{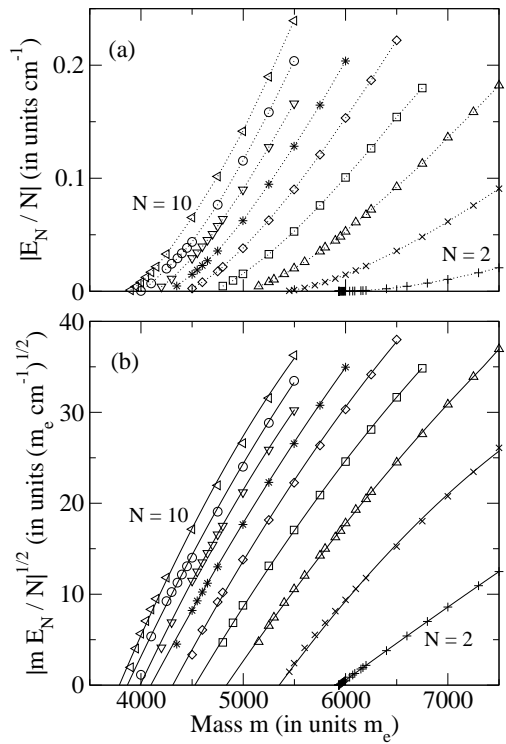}
\caption{
(a) Symbols show the absolute value of the numerically determined 
ground state energies $E_N/N$ per particle
as a function of the atom mass $m$ for $N=2-10$; the
energies for $N=2$ and $3$ are also shown
in Figs.~\protect\ref{fig3} and \protect\ref{figtrimer},
respectively.
To guide the eye, dotted lines connect energies for the same $N$.
(b) Symbols show the scaled energies $\sqrt{m |E_N| /N}$
as a function of the atom mass for $N=2-10$.
Solid lines show our fits to the scaled energies treating 
$c_1^{(N)}$, $c_2^{(N)}$ and $m^{(N)}_*$ as fitting parameters.
For each $N$, the
crossing point of the fit with the zero-energy 
line predicts the critical mass
$m_*^{(N)}$. 
}
\label{fig4}
\end{figure}

We now turn to the discussion of weakly-bound clusters 
with up to $N=40$ atoms. Symbols in Fig.~\ref{fig4}(a)
show the absolute value of
the ground state energies $E_N/N$ per particle
as a function of the atom mass $m$
for $N=2-10$. 
The energies for $N=2$ and $3$ 
are also shown
in Figs.~\ref{fig3} and \ref{figtrimer}, respectively.
The statistical uncertainties of the DMC energies $E_N/N$
per particle, $N \ge 4$,
are not shown in Fig.~\ref{fig4} since they are smaller than the symbol 
sizes. The 
overall behavior of the ground state energies
is similar for all $N$. 
Below, we use our energies for $N=2-10$ (see Fig.~\ref{fig4}), 
and $N=20$ and 40 (not shown)
to determine the critical masses $m_*^{(N)}$.

Symbols in Fig.~\ref{fig4}(b) show the scaled
energies $\sqrt{m |E_N|/N}$ for $N=2-10$ as a function of the
atom mass $m$.
To determine the critical masses $m_*^{(N)}$, we fit the scaled energies
$\sqrt{m |E_N|}$
for each $N$ to the functional form given in Eq.~(\ref{eq_expand})
with $c_i^{(2)}$ ($i \le 2$) 
and $m_*^{(2)}$ replaced by $c_i^{(N)}$ and $m_*^{(N)}$,
respectively.
The resulting fitting parameters $c_1^{(N)}$, $c_2^{(N)}$ and $m_*^{(N)}$
are given in Table~\ref{tab1};
$c_1^{(N)}$ 
increases with increasing $N$,
and $m_*^{(N)}$ decreases with increasing $N$.
The fitting parameter $c_2^{(N)}$ is negative for all $N$
considered; 
its largest uncertainty 
is about 40\% for $N=20$.
As discussed in Sec.~\ref{sec_trimer},
the critical masses $m_*^{(N)}$ predicted by our
fits are expected 
to be lower bounds to the exact threshold value. An upper bound 
is given by the smallest mass for which we report a bound state.
A more precise extrapolation of the threshold value is complicated by the
fact that the DMC energies have error bars and that the determination
of the energy becomes numerically more demanding the closer
the system's mass is
to the critical mass $m_*^{(N)}$.

\begin{figure}
      \centering
      \includegraphics[angle=0,width=8cm]{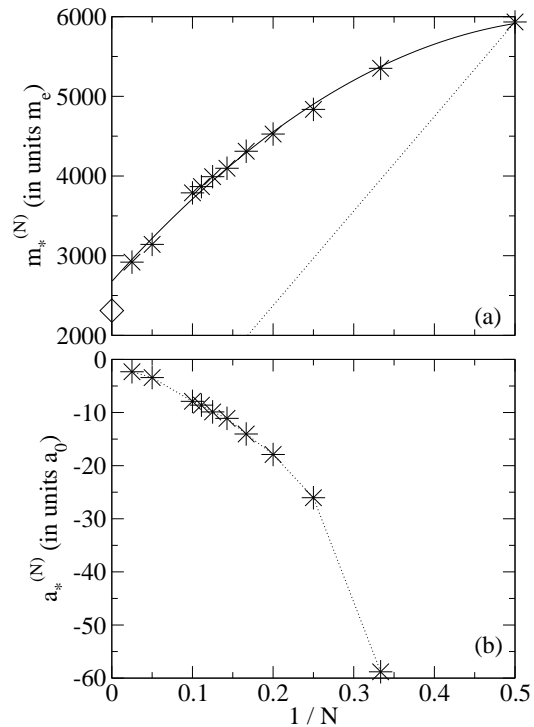}
\caption{
(a)
Asterisks show the critical masses $m^{(N)}_*$, predicted by
our fits to the scaled ground state energies $\sqrt{m |E_N|}$,
as a function of 
$1/N$ for $N=2-10$, $20$ and 40.
The diamond shows an upper bound for the critical mass $m^{(\infty)}_*$
of the bulk system and
the solid line shows our three-parameter fit, Eq.~(\ref{eq_critmass}), with
$D = 2676(47)m_e$, $E=11325(458) m_e$ and $F=-9696(852) m_e$
to the critical masses 
$m_*^{(N)}$ 
(the numbers in round brackets denote the uncertainty of the fit,
neglecting possible uncertainties of the critical masses).
The dotted line shows a lower bound for $m_*^{(N)}$
using the equal sign in the relationship $m^{(N)}_* \ge m_*^{(N-1)} (N-1)/N$
and $m_*^{(2)}=5933.4m_e$.
As expected, our numerically determined critical masses
lie above this analytical bound.
(b) Asterisks show the critical scattering length $a_*^{(N)}$
as a function of $1/N$ for $N=2-10$, $20$ and 40.
To guide the eye, a dotted line connects the symbols.
}
\label{fig5}
\end{figure}

Asterisks in Fig.~\ref{fig5}(a) 
show the critical masses $m_*^{(N)}$ predicted by
our fits for $N=2-10$, $20$ and 40
as a function of $1/N$. 
A diamond in Fig.~\ref{fig5}(a)
shows the upper bound for the critical mass $m_*^{(\infty)}$ of the
bulk system, i.e.,
$m_*^{(\infty)}=2311.0m_e$
(see Sec.~\ref{sec_system}). 
We choose the $1/N$-scale since it allows the critical mass
for the dimer and the bulk system to be shown on the same graph;
to the best of our knowledge, the functional dependence of
$m_*^{(N)}$ on the system size is unknown.
Our critical mass for $N=40$ is significantly larger than the upper 
bound $m_*^{(\infty)}$ determined variationally for the bulk system.
Indeed, a three-parameter fit of the form
\begin{eqnarray}
\label{eq_critmass}
m_*^{(N)} = D + E / N + F / N^2
\end{eqnarray}
to our critical masses 
for up to 40 atoms, shown by a solid line in Fig.~\ref{fig5}(a), 
predicts a larger critical mass for the bulk system than
the upper bound $m_*^{(\infty)}=2311.0m_e$ at which the
scattering length crosses zero.
We speculate that our calculations for comparatively small $N$ 
cannot be used to extrapolate $m_*^{(\infty)}$ reliably since
the ratio of bulk to surface atoms increases appreciably
with increasing $N$.
Furthermore, our extrapolated critical 
masses have non-negligible uncertainties.
To connect the results discussed here to a realistic
physical system,
we note that 
homogeneous atomic spin-polarized hydrogen, 
which has an atom mass of $m=1837m_e$ and interacts through
a sum of two-body potentials only slightly different
from that considered here, exists under normal pressure 
as a gas and not as a liquid~\cite{ette75}.

The critical mass $m_*^{(N)}$ of the $N$-body system is bounded
by the critical mass of the system with $N-1$ particles through
$m_*^{(N)} \ge m_*^{(N-1)} (N-1)/N$~\cite{jens04}. A dotted line in
Fig.~\ref{fig5}(a) shows this lower bound,
assuming $m_*^{(2)}=5933.4m_e$.
Figure~\ref{fig5}(b) indicates that this analytical estimate 
provides a weak bound for all bosonic systems considered here.
An upper bound is given by $m_*^{(N)} = m_*^{(N-1)}$.

To relate our critical masses to an experimentally tunable
parameter, 
we calculate the
scattering length for each $m_*^{(N)}$ and refer to it
as the critical scattering length $a_*^{(N)}$.
Asterisks in Fig.~\ref{fig5}(b) show 
the critical scattering length $a_*^{(N)}$ 
as a function of $1/N$.
Figure~\ref{fig5}(b) suggests that
Borromean trimers exist for $a \le -58.8 a_0$
and Borromean tetramers for $a \in [ -58.8 a_0, -26.0 a_0]$.
Investigation of the stability of these weakly-bound
Borromean states is beyond the scope of this paper.

\subsection{Correlations}
We now investigate correlations between 
energies of the three- and four-particle systems.
The description of universal
properties of the trimer~\cite{braa06}, 
such as the description of Efimov 
states, requires two
parameters, 
a two-body momentum scale $\mu^{(2)}$
(typically taken to be inversely proportional to
the $s$-wave scattering length $a$) and a three-body 
momentum scale $\mu^{(3)}$ (in some studies, the
three-body parameter $\Lambda_*$ is used instead~\cite{beda99,beda99a}).
While the universal behaviors of the trimer are quite well 
understood~\cite{braa06}, 
much less is known about those of larger systems.
For example, although one expects a new momentum scale $\mu^{(4)}$ to be needed
for the description of universal properties of the 
tetramer~\cite{adhi95,yama06}, 
there is evidence
that at least some 
observables of the tetramer near threshold are independent
of this new momentum scale~\cite{plat04}.
In this context, a number of studies have focused on the Tjon 
line~\cite{tjon75,tjon81,naka78,naka79,lim80},
which was first investigated
in nuclear physics and refers to the approximately linear correlation
between the energies of the four-nucleon and the
three-nucleon system~\cite{tjon75,tjon81}. 
In the following, correlations between 
our trimer and tetramer energies,
which are calculated for realistic atom-atom interactions,
are demonstrated.

\begin{figure}
      \centering
      \includegraphics[angle=0,width=8cm]{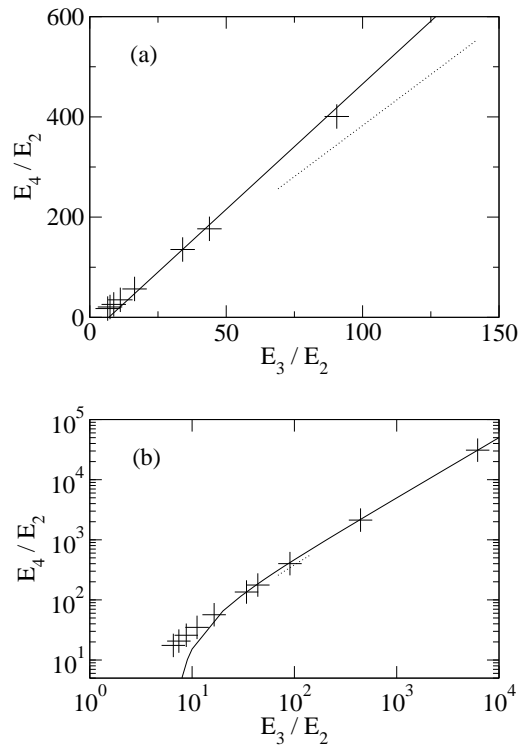}
\caption{
Pluses show the energy ratio $E_4/E_2$ as a function of the energy ratio
$E_3/E_2$ ($E_4$, $E_3$ and $E_2$ denote ground state energies)
on (a) a linear scale and (b) a log-log scale.
Solid lines show our two-parameter fit (see text).
Dotted lines show the result obtained within an
effective quantum mechanical approach~\protect\cite{plat04}. 
The data with the largest energy ratios correspond to systems 
closest to threshold, and those with the smallest energy ratios 
correspond to
systems farthest away from threshold.
}
\label{fig6}
\end{figure}

Pluses in Figs.~\ref{fig6}(a) and (b) show the ratio 
between the ground state energies
$E_4$ and $E_2$ 
as a function of the ratio between the ground state energies
$E_3$ and $E_2$ on a linear and 
double-logarithmic scale, respectively. 
Unlike in Tjon's original work for fixed dimer energy~\cite{tjon75,tjon81}, 
we scale the trimer and tetramer energies in Fig.~\ref{fig6}
by the dimer energy since $E_2$ depends on
the atom mass $m$.
For each data point,
the ground state
energies $E_2$, $E_3$ and $E_4$ are calculated for the same atom mass $m$.
The systems closest to threshold are those with the largest energy ratios.
For the smallest mass, $m=5950 m_e$, included in Fig.~\ref{fig6}(b),
the absolute value of $E_2$ is
nearly four orders of magnitude smaller than that of $E_3$,
and more than four orders of magnitude smaller than that of $E_4$.
A two-parameter fit of the form
$E_4/E_2= B_3 + C_3 E_3/E_2$, 
shown by solid lines in Figs.~\ref{fig6}(a) and (b),
describes the dependence of $E_4/E_2$ on $E_3/E_2$ quite well
(especially for systems close to threshold),
thus confirming the existence of the Tjon line
for atomic clusters. In particular, we find
$B_3 = -34.9(8.3)$ and $C_3= 5.008 (5)$ (see also
Table~\ref{tabfit}), where the numbers
in brackets indicate the uncertainty of the fit, neglecting possible
inaccuracies of the energy ratios.

For comparison, dotted lines in Figs.~\ref{fig6}(a) and (b)
show a result derived within an effective
quantum mechanics approach applied to bosonic
clusters with $N=2-4$ helium atoms~\cite{plat04}. This study 
finds $B_3=-24.752$
and $C_3=4.075$ for $69 \le E_3/E_2 \le 142$.
This range of $E_3/E_2$ values is significantly smaller
than that considered in the present paper [solid lines in
Figs.~\ref{fig6}(a) and (b)].
The slope of the Tjon line derived within
the effective quantum mechanical approach, applied to helium clusters, 
is somewhat smaller than our slope, which is
derived from a series of numerical calculations for
weakly-bound Van der Waals clusters.
We find that our slope decreases if we perform a fit
that excludes energy ratios for systems very close to threshold.
This may explain the discrepancy between the results
obtained within the two approaches.

It has been argued that, 
if the four-body momentum scale $\mu^{(4)}$ coincides with
the three-body momentum scale $\mu^{(3)}$~\cite{yama06},
the  slope of the Tjon line is about five. This argument
suggests that the systems studied in the present paper have approximately
equal three- and four-body momentum scales.
It has further been suggested that the existence of equal three- and
four-body momentum scales can be traced back to the repulsive core
of the two-body interactions~\cite{yama06}.
This interpretation suggests that the Tjon line is roughly five
for all atomic systems near threshold and 
that the
ground state energy of any weakly-bound tetramer interacting 
additively through Van der Waals potentials with repulsive core 
can be estimated quite reliably if the corresponding dimer and trimer 
ground state energies are known. 
This interpretation could be checked more rigorously by performing a series
of calculations
for systems interacting additively through
a shape-dependent two-body potential, which depends on
a parameter that controls the softness of the repulsive short-range
part of the potential.
For a ``soft'' repulsive core, the four-body momentum scale
should deviate from the three-body scale and 
the slope of the Tjon line should deviate from five.
Such a study is
beyond the scope of this paper.

\begin{figure}
      \centering
      \includegraphics[angle=270,width=8cm]{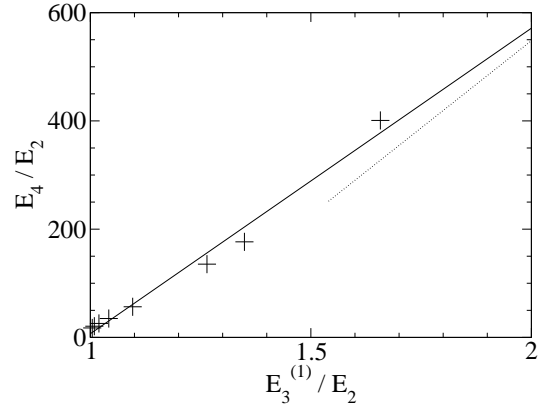}
\caption{
Pluses show the energy ratio $E_4/E_2$ as a function of the energy ratio
$E_3^{(1)}/E_2$ on a linear scale ($E_2$ and $E_4$
denote ground state energies, and $E_3^{(1)}$ denotes the first
excited state energy with $J=0$).
A solid line shows our two-parameter fit (see text).
A dotted line shows the results obtained within an
effective quantum mechanical approach~\protect\cite{plat04}. 
The data with the largest energy ratios correspond to systems 
closest to threshold, and those with the smallest energy ratios correspond to
systems farthest away from threshold.
}
\label{fig6exc}
\end{figure}

We now consider correlations between the 
tetramer ground state energy $E_4$ and the trimer excited state
energy $E_3^{(1)}$, both scaled
by the dimer ground state energy $E_2$.
Pluses in Fig.~\ref{fig6exc} show the energy ratio $E_4/E_2$
as a function of the energy ratio $E_3^{(1)}/E_2$ 
on a linear scale. A solid line shows our two-parameter fit 
$E_4/E_2 = \bar{B}_3 + \bar{C}_3 E_3^{(1)}/E_2$ 
with $\bar{B}_3 = -558(35)$ and $\bar{C}_3 =565(29)$, while
a dotted line shows that derived in Ref.~\cite{plat04} for 
$1.54 \le E_3^{(1)}/E_2 \le 2$ with 
$\bar{B}_3=-742.0$ and $\bar{C}_3=645.1$.
The agreement of the slopes, derived within  two different 
frameworks and applied to two different systems, is quite
reasonable. We now use the linear dependence of $E_4/E_2$
on $E_3/E_2$ and of $E_4/E_2$
on $E_3^{(1)}/E_2$ to predict the slope
$\tilde{C}_3$ for the linear dependence of  $E_3/E_2$
on $E_3^{(1)}/E_2$,
$E_3/E_2= \tilde{B}_3+ \tilde{C}_3 E_3^{(1)}/E_2$.
From our slopes $C_3$ and $\bar{C}_3$, we obtain $\tilde{C}_3=112(6)$. 
For comparison, a fit to our data 
gives $ \tilde{C}_3=124(5)$. 
The good agreement lends support to the predictive power of the 
approximately linear correlations between energy ratios
of clusters with varying number of atoms.

\begin{figure}
      \centering
      \includegraphics[angle=0,width=9cm]{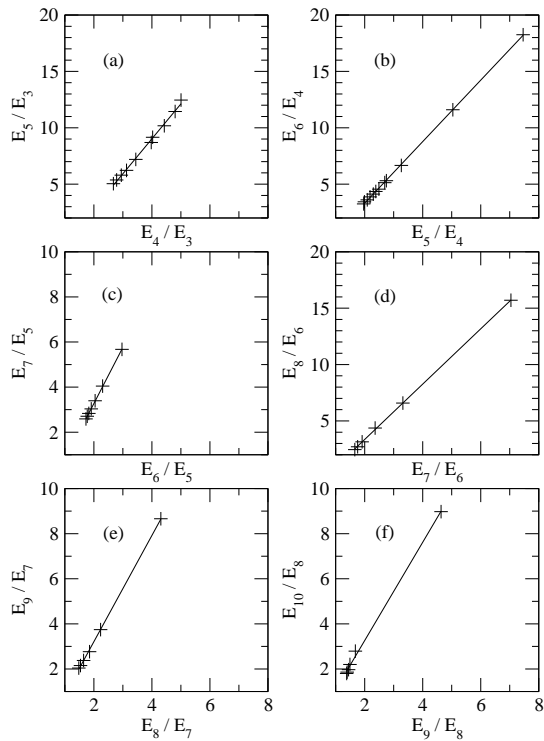}
\caption{
Pluses show the energy ratio $E_{N+1}/E_{N-1}$
as a function of the energy ratio $E_{N}/E_{N-1}$
for (a) $N=4$,
(b) $N=5$,
(c) $N=6$,
(d) $N=7$,
(e) $N=8$, and
(f) $N=9$. Solid lines show linear fits of the form
$E_{N+1}/E_{N-1}= B_{N} + C_{N} E_{N}/E_{N-1}$ (see text).
Note that the range of the
vertical axis 
extends from 1 to 8 in all panels while that of the horizontal axis 
varies.}
\label{fig7}
\end{figure}

Correlations between the 
energies of two clusters differing in size
by one atom, i.e., a linear relationship
of the form $E_{N+1} = b_N + c_N E_N$,
have been predicted analytically based
on a separable approximation scheme for any cluster size~\cite{pern79}
and numerically by performing variational calculations
for small mixed $^3$He$_i$-$^4$He$_j$ clusters
with $i+j \le 5$~\cite{naka79}.
We find that our energies of clusters differing in size by one atom
are not well described by such a linear two-parameter fit.
If we instead scale, 
as in the investigation of the correlations between the tetramer
and
trimer energies, 
 our 
ground state 
energies of the $(N+1)$-
and $N$-atom clusters by the energy of the $(N-1)$-atom cluster,
we find 
an approximately linear relationship.
Pluses in Figs.~\ref{fig7}(a)-(f) show the energy ratios
$E_{N+1}/E_{N-1}$ 
as a function of $E_{N}/E_{N-1}$
for $N=4-9$, 
and solid lines a linear fit 
of the form
$E_{N+1}/E_{N-1}=B_{N} + C_{N} E_{N}/E_{N-1}$. The fitting parameters 
$B_{N}$ and $C_{N}$ are summarized in Table~\ref{tabfit}.
\begin{table}
\begin{tabular}{l|ll} 
$N$ & $B_{N}$ & $C_{N}$ \\ \hline
  3	& $-34.9(8.3)$  & $5.008(5)$  \\
  4	& $-3.37(27)$ &	$3.10(8)$  \\
  5	& $-2.27(08)$ &	$2.78(2)$  \\
  6	& $-1.71(03)$ &	$2.49(2)$  \\
  7	& $-1.55(07)$ &	$2.45(2)$  \\
  8	& $-1.50(06)$ &	$2.36(3)$  \\
  9	& $-1.11(11)$ & $2.18(5)$	  
\end{tabular}
\caption{
Fitting parameters $B_{N}$ and $C_{N}$ for $N=3-9$.
Numbers in brackets denote the uncertainty of the two-parameter fit
$E_{N+1}/E_{N-1} = B_N + C_N E_N / E_{N-1} $,
neglecting possible uncertainties of the energy ratios.
}
\label{tabfit}
\end{table}
We refer to the approximately linear dependence of 
the energy ratios of
clusters, which is 
illustrated in Fig.~\ref{fig7}, as generalized Tjon lines.
It will be interesting to investigate the implications
of the behaviors of these generalized Tjon lines for 
the universal properties of weakly-bound 
bosonic clusters.

\section{Structural Properties}
\label{sec_results_structure}
\begin{figure}
      \centering
      \includegraphics[angle=270,width=8cm]{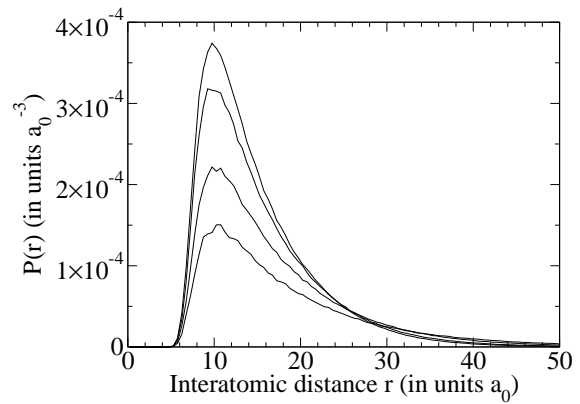}
\caption{
Pair distribution functions $P(r)$ as a function of the interparticle
distance
$r$ 
for $N=4$ and four different masses, i.e.,
$m=5950m_e$, $5750m_e$, $5400m_e$ 
and $5150m_e$ (from top to bottom).
The 
pair distribution functions
are calculated using the MC
estimator given in Eq.~(\ref{eq_extra}), which combines
the VMC and DMC expectation values.
Statistical uncertainties are not shown for clarity.
}
\label{fig9}
\end{figure}

This section presents selected structural properties
of weakly-bound
bosonic clusters in their ground state.
The expectation values for $N \ge 4$ are 
calculated using the MC estimator given in Eq.~(\ref{eq_extra}).
Figure~\ref{fig9} shows the pair distribution function $P(r)$
for $N=4$ and
four different masses, i.e., $m=5950m_e$, $5750m_e$, $5400m_e$
and $5150m_e$. 
The pair distribution function $P(r)$ 
indicates the likelihood of finding two particles at a
distance $r$ from each other and is normalized so that
\begin{eqnarray}
\int_0^{\infty} P(r) \, r^2 dr = 1 .
\end{eqnarray}
As the mass decreases, the maximum of $P(r)$, which is located
at $r \approx 10 a_0$, decreases. Furthermore, the pair distribution functions
extend to significantly larger $r$ values for small $m$ than for large
$m$. For example,
the largest interparticle distance sampled in our DMC runs for $m=5950m_e$
is $r \approx 100 a_0$ while that for $m=5150 m_e$ is 
$r \approx 200 a_0$.
We find that the densities, not shown, 
of the weakly-bound clusters studied in this
paper 
are structureless
and
do not possess any shell structure. The highly-diffuse
clusters can thus be most appropriately
thought of as ``diffuse liquid blobs''.

\begin{figure}
      \centering
      \includegraphics[angle=270,width=8cm]{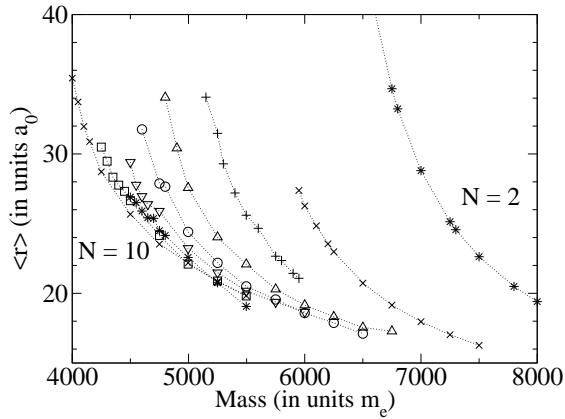}
\caption{
Expectation value $\langle r \rangle _N$ of the interparticle distance 
for clusters in the ground state with $N=2-10$ (from right to left)
as a function of the atom mass $m$.
The expectation values for $N \ge 4$ 
are calculated using the extrapolated
estimator, Eq.~(\ref{eq_extra}), which combines
the VMC and DMC expectation values.
Errorbars, not shown, are at most about three times as
large as the symbol sizes.
}
\label{fig10}
\end{figure}

To compare the structural behaviors of clusters with 
different $N$, symbols in Fig.~\ref{fig10}
show the expectation values of the interparticle 
distance $r$ for clusters in the ground state with $N=2-10$ 
(denoted by $\langle r \rangle _N$)
as a function of
the atom mass $m$.
For fixed $N$, $\langle r \rangle _N$ decreases,  as expected,
with decreasing mass 
$m$.
For a given mass, 
$\langle r \rangle_N$ decreases with increasing
$N$. This behavior is consistent with the fact that the energy 
per particle for fixed mass decreases with increasing $N$.
Furthermore, for fixed $m$, the expectation values $\langle r \rangle _N$
should reach a constant in the large $N$ limit.
Indeed, Fig.~\ref{fig10} indicates that the difference between 
$\langle r \rangle _N$
for two clusters differing in size by one atom is smaller for large
than for small $N$.

\begin{figure}
      \centering
      \includegraphics[angle=270,width=8cm]{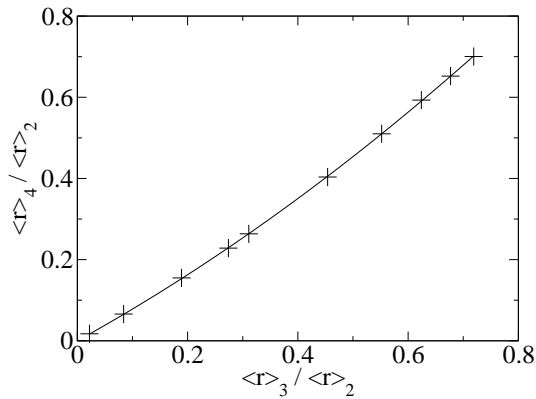}
\caption{
Ratio $\langle r \rangle_4 / \langle r \rangle_2$
as a function of the ratio
$\langle r \rangle_3 / \langle r \rangle_2$.
Pluses show our numerically determined ratios while the solid line
shows a two-parameter fit (see text). 
}
\label{fig11}
\end{figure}

It has been suggested that scaling functions, which 
allow the structural properties of the tetramer
to be expressed in terms of expectation values of the dimer and trimer,
exist~\cite{yama06}; the exact functional forms are, however, 
to the best of our knowledge unknown.
Symbols in Fig.~\ref{fig11} show the ratio 
$\langle r \rangle_4 / \langle r \rangle_2$ 
as a function
of the ratio 
$\langle r \rangle_3 / \langle r \rangle_2$. 
For each data point, the expectation values $\langle r \rangle _4$,
$\langle r \rangle _3$ and $\langle r \rangle _2$ are
calculated for the same mass.
The errorbars, not shown, are smaller than twice the
size of the symbols. 
These ratios are well described by a two-parameter fit of the form
$\langle r \rangle_4 /\langle r \rangle_2 =
G \langle r \rangle_3 /\langle r \rangle_2 + 
H  (\langle r \rangle_3 /\langle r \rangle_2)^2 $ with
$G=0.751(5)$ and $H = 0.313(7)$ (solid line).
We hope that the structural properties presented here will stimulate and
aid further studies of weakly-bound bosonic clusters.

\section{Conclusion}
\label{sec_conclusion}
The physics of weakly-bound few-body systems can experimentally be
investigated using ultracold atomic gases. Indeed, the
observation of resonances in an ultracold Bose gas has recently
been interpreted as evidence for the presence of loosely-bound
Efimov trimers~\cite{krae06}. Resonances associated with
tetramers have also been reported~\cite{chin05}.
These experiments may just be the beginning of detailed studies
of the rich behaviors
of few-body systems under controlled conditions.
On the theoretical side, little is known about the
universal near-threshold
behaviors of three-dimensional
systems with more than three particles. The reason is
that analytical treatments become increasingly more complex as the number
of degrees of freedom increases. On the other hand, numerical treatments 
are complicated by the fact that the kinetic and potential energy
nearly cancel, thus requiring both to be calculated with high accuracy.

This paper presents a detailed study of the near-threshold
behaviors of weakly-bound three-dimensional bosonic 
clusters with up to $40$ atoms, 
for which the underlying potential energy surface
is written as a sum of realistic Van der Waals atom-atom potentials
with short-range repulsion and attractive long-range tail.
In particular, we determine the critical mass $m_*^{(N)}$
for clusters with $N=2-10$, $20$ and 40 by performing calculations
for each cluster as a function of the atom mass $m$.
To the best of
our knowledge, these are the first calculations
that attempt an accurate determination of the critical coupling
strengths of clusters with up to 40 atoms.
Our critical masses are compared to analytical bounds.
Furthermore, we show that our numerically determined 
three- and four-particle energies, scaled by the corresponding
dimer energies, fall on the Tjon line.
We present numerical evidence that the scaled energies of 
larger clusters differing
in size by one atom 
also correlate linearly, i.e., the energy ratios fall on what we refer to as
generalized Tjon lines. 
Motivated by a recent calculation based on zero-range
models~\cite{yama06}, we 
speculate that all atomic cluster systems show similar 
near-threshold behaviors.
Finally, we present selected structural properties of weakly-bound
few-body systems.

In closing, we emphasize that the near-threshold behavior
of clusters crucially depends on the dimensionality.
For example, the near-threshold behavior of weakly-bound two-dimensional
few-body 
systems~\cite{hamm04,blum05} 
is very different from that presented here for three-dimensional
systems.
We hope that our work will stimulate further 
experimental and theoretical work
on weakly-bound clusters.

We thank Chris Greene for fruitful discussions.
Acknowledgement is made to the Donors of The
Petroleum Research
Fund, administered by the American Chemical Society, 
and the NSF (grant ITR-0218643) for support of this research.

\end{document}